\documentclass[smallextended]{svjour3}

\smartqed

\usepackage{graphicx}

\journalname{General Relativity and Gravitation}

\begin{document}

\title{Vacuum for a massless quantum scalar field outside a collapsing shell in anti-de Sitter space-time}

\titlerunning{Vacuum for a scalar field outside a collapsing shell in adS}

\author {Paul G.~Abel \and Elizabeth Winstanley}

\institute{
P.~G.~Abel \at
Centre for Interdisciplinary Science,
University of Leicester,
Leicester. LE1 7RH \\
United Kingdom.
\email{pga3@leicester.ac.uk}
\and
E.~Winstanley \at
Consortium for Fundamental Physics,
School of Mathematics and Statistics, \\
The University of Sheffield,
Hicks Building,
Hounsfield Road,
Sheffield. S3 7RH \\
United Kingdom.
\email{E.Winstanley@sheffield.ac.uk}
}

\date{\today}

\maketitle

\begin{abstract}
We consider a massless quantum scalar field on a two-dimensional space-time describing a thin shell of matter collapsing to form a Schwarzschild-anti-de Sitter black hole. At early times, before the shell starts to collapse, the quantum field is in the vacuum state, corresponding to the Boulware vacuum on an eternal black hole space-time.
The scalar field satisfies reflecting boundary conditions on the anti-de Sitter boundary.
Using the Davies-Fulling-Unruh prescription for computing the renormalized expectation value of the stress-energy tensor, we find that at late times the
black hole is in thermal equilibrium with a heat bath at the Hawking temperature, so the quantum field is in a state analogous to the Hartle-Hawking vacuum on an eternal black hole space-time.
\end{abstract}

\section{Introduction}
\label{sec:intro}

The study of quantum field theory on black hole space-times has a long history, including the discovery of Hawking radiation from asymptotically flat black holes \cite{Hawking:1974sw}.
At late times, far from an asymptotically flat black hole formed by gravitational collapse, a static observer sees an outgoing flux of thermal radiation at the Hawking temperature.
Hawking's original derivation \cite{Hawking:1974sw} considered the fully dynamical space-time consisting of matter collapsing to form a black hole.
However, quantum field theory on static space-times is technically much easier than on dynamical space-times.
For an eternal Schwarzschild black hole, the Unruh state \cite{Unruh:1976db} contains an outgoing thermal flux of radiation at future null infinity, and therefore is the state analogous to that at late times for a black hole formed by gravitational collapse.

Since all two-dimensional space-times are locally conformally flat, quantum field theory on two-dimensional space-times is rather simpler than in four space-time dimensions, particularly if one is considering dynamical space-times.
In particular, two-dimensional black hole space-times have proven to be useful toy models for understanding various aspects of quantum field theory on four-dimensional black holes, in particular Hawking radiation and black hole evaporation \cite{Fabbri:2005mw}.
One advantage of working in two space-time dimensions is that there is an exact prescription \cite{Davies:1976ei} for computing the renormalized expectation value of the stress-energy tensor $\langle T_{\mu \nu } \rangle $ for a massless scalar field.
Applying this prescription to a two-dimensional asymptotically flat space-time describing a black hole formed by gravitational collapse, it is found that at late times after the gravitational collapse, far from the black hole $\langle T_{\mu \nu }\rangle $ contains an outgoing flux of energy compared to that at early times before the collapse starts \cite{Davies:1976ei}.
This outgoing flux is the Hawking radiation and is also found in the expectation value $\langle T_{\mu \nu } \rangle $ for the Unruh vacuum state on an eternal two-dimensional Schwarzschild black hole \cite{Christensen:1977jc}.

A natural question is how the above picture is modified in space-times with different asymptotic structure.
The case of a Schwarzschild-de Sitter black hole was studied many years ago.
Working in two space-time dimensions, any static quantum state defined in the region between the event and cosmological horizons must have a stress-energy tensor $\langle T_{\mu \nu } \rangle $ which diverges at either the event or the cosmological horizon
\cite{Hiscock:1989yw,Denardo:1980na}, in agreement with the four-dimensional Kay-Wald theorem that there is no stationary, nonsingular state on Schwarzschild-de Sitter space-time \cite{Kay:1988mu}.
However, a nonstatic quantum state can be defined which is regular on both the event and cosmological horizons of an eternal Schwarzschild-de Sitter black hole \cite{Markovic:1991ua,Tadaki:1990aa}.
The corresponding state for a two-dimensional Schwarzschild-de Sitter black hole formed by gravitational collapse has also been considered \cite{Tadaki:1990cg}.

A great deal of attention has been given to Hawking radiation from black holes in asymptotically anti-de Sitter (adS) space-time,
particularly in the context of the adS/CFT (conformal field theory) correspondence (see, for example, \cite{Aharony:1999ti} for a review).
Various authors have considered the definition of states for quantum fields on asymptotically adS black hole backgrounds, largely focussing on the analogues of the Boulware \cite{Boulware:1974dm} and Hartle-Hawking \cite{Hartle:1976tp} states.
Some work on these states for eternal Schwarzschild-anti-de Sitter (SadS) black holes in two or more dimensions can be found in \cite{Maldacena:2001kr,Spradlin:1999bn,Morita:2008qn,Solodukhin:1995te,Hubeny:2009rc,Ortiz:2011mi,Ng:2014kha}.

In this paper we address the question of the existence, for asymptotically adS black holes, of a quantum state analogous to the Unruh vacuum \cite{Unruh:1976db} on asymptotically flat black hole space-times.
Working in two space-time dimensions for simplicity,  we consider an SadS black hole formed by the gravitational collapse of a thin shell of matter in anti-de Sitter space-time.
This scenario has been studied by a number of authors (see, for example,
\cite{Hemming:2000as,Danielsson:1999zt,Danielsson:1999fa,Giddings:2001ii}),
largely (but not exclusively) from the point of view of adS/CFT and with a particular focus on the boundary theory.
Here we work entirely in the bulk space-time, using the standard techniques of quantum field theory in curved space. We follow the approach of \cite{Davies:1976ei,Markovic:1991ua,Tadaki:1990cg} to compute the renormalized expectation value of the stress-energy tensor for a massless quantum scalar field propagating on this background.
Our aim is to compute the difference between the expectation values at early times, before the collapse starts, and late times, long after the black hole has formed, comparing the result with that in \cite{Davies:1976ei} for the asymptotically flat case.

The outline of this paper is as follows.
In section \ref{sec:DFU} we briefly review the method of Davies-Fulling-Unruh (DFU) \cite{Davies:1976ei} for computing the renormalized expectation value of the stress-energy tensor on a two-dimensional space-time.
Our collapse scenario is discussed in section \ref{sec:collapse}, focussing on the definition of particular ingoing and outgoing null coordinates which are required for the DFU method.
We then apply the DFU method in section \ref{sec:Tmunu} to compute the difference in the expectation values of the stress-energy tensor at early and late times.
Section \ref{sec:conc} contains further discussion and our conclusions.
Throughout this paper we use units in which $c=\hbar =G=k_{B}=1$, and our two-dimensional metric signature is $(-,+)$.

\section{Computation of $\langle T_{\mu \nu } \rangle $ in two-dimensional space-times}
\label{sec:DFU}

We now briefly review the Davies-Fulling-Unruh (DFU) method \cite{Davies:1976ei} to compute the renormalized expectation value of the stress-energy tensor, $\langle T_{\mu \nu } \rangle $, for a massless quantum scalar field on a general two-dimensional space-time.

Any two-dimensional space-time is locally conformally flat and therefore the metric can be written in the following form:
\begin{equation}
ds^{2} = - C(u,v) \, du \, dv ,
\label{eq:genmetric}
\end{equation}
where $C(u,v)$ is a function of the null coordinates $u$ and $v$.
A change of null coordinates of the form ${\bar {u}}={\bar {u}}(u)$, ${\bar {v}}={\bar {v}}(v)$ does not change the form (\ref{eq:genmetric}), but does change the overall function $C$:
\begin{equation}
ds^{2} = -{\bar {C}}({\bar {u}}, {\bar {v}} ) \, d{\bar {u}} \, d{\bar {v}} ,
\label{eq:metric}
\end{equation}
where
\begin{equation}
{\bar {C}}({\bar {u}}, {\bar {v}}) = C(u,v) \frac {du}{d{\bar {u}}} \frac {dv}{d{\bar {v}}} .
\end{equation}

We now consider a massless scalar field satisfying the Klein-Gordon equation on the space-time with metric (\ref{eq:metric}). Following \cite{Davies:1976ei}, we assume that there exist null coordinates ${\bar {u}}$ and ${\bar {v}}$ such that the outgoing and ingoing wave modes of the scalar field take, respectively, the standard form
\begin{equation}
\frac {1}{{\sqrt {4\pi \left| \omega \right| }}} e^{-i\omega {\bar {u}}} , \qquad
\frac {1}{{\sqrt {4\pi \left| \omega \right| }}} e^{-i\omega {\bar {v}}} ,
\label{eq:DFUmodes}
\end{equation}
throughout the space-time.
For a general dynamical space-time, the definition of suitable null coordinates $({\bar {u}},{\bar {v}})$ is nontrivial.
The metric function ${\bar {C}}$ (\ref{eq:metric}) will also, in general, be a complicated function of $({\bar {u}},{\bar {v}})$, particularly for a dynamical space-time, in which case all the time-dependence is included in ${\bar {C}}$.

Positive frequency modes are defined to be modes such that $\omega >0$, and the scalar field is written as a sum over positive and negative frequency modes.
The field is quantized by promoting the expansion coefficients to operators.
We define the vacuum state $| 0 \rangle $ to be that state annihilated by all the operators which are coefficients of positive frequency modes.

In \cite{Davies:1976ei} it is shown that the renormalized stress-energy tensor for the quantum scalar field in the state $| 0 \rangle $ takes the following form in $({\bar {u}}, {\bar {v}})$ coordinates (see also \cite{Birrell:1982ix}):
\begin{equation}
\langle T_{\mu }^{\nu } \rangle =
\frac {1}{{\sqrt {-g}}} \langle T_{\mu }^{\nu } \left[ \eta  \right] \rangle
 + \theta _{\mu }^{\nu }
 - \frac {{\mathcal {R}}}{48\pi } \delta _{\mu }^{\nu }.
 \label{eq:DFU}
\end{equation}
Here the metric tensor $g_{\mu \nu }$  (\ref{eq:metric}) has determinant $g$, and ${\mathcal {R}}$ is the Ricci scalar of the metric (\ref{eq:metric}).

The first term in (\ref{eq:DFU}), $\langle T_{\mu }^{\nu } \left[ \eta \right] \rangle $, is a nonlocal contribution from flat space-time and depends on the ranges of the null coordinates $({\bar {u}},{\bar {v}})$ in (\ref{eq:metric}).
If the space-time with metric (\ref{eq:metric}) is conformal to the whole of Minkowski space-time, then ${\bar {u}}$ and ${\bar {v}}$ may take any real value and $\langle T_{\mu }^{\nu } \left[ \eta \right] \rangle =0$.
If, on the other hand, the space-time under consideration is conformal only to part of Minkowski space-time (so that the ranges of ${\bar {u}}$ and ${\bar {v}}$ are constrained), then $\langle T_{\mu }^{\nu } \left[ \eta \right]
\rangle $ is equal to the vacuum expectation value of the massless scalar field on that portion of Minkowski space-time defined by the coordinates $({\bar {u}},{\bar {v}})$.
If we are considering a dynamical space-time, then all the time-dependence is contained in the complicated metric function ${\bar {C}}$ in (\ref{eq:metric}).
By definition the values taken by the null coordinates $({\bar {u}}, {\bar {v}})$ do not depend on time and therefore $\langle T_{\mu }^{\nu } \left[ \eta \right] \rangle $ also does not depend on time.

The second term in (\ref{eq:DFU}), $\theta _{\mu \nu }$, is also nonlocal, and has components \cite{Davies:1976ei}:
\begin{equation}
\theta _{{\bar {u}}{\bar {u}}}  =
- \frac {1}{12\pi } {\bar {C}}^{\frac {1}{2}} \left( {\bar {C}}^{-\frac {1}{2}} \right) _{,{\bar {u}}{\bar {u}}}
\qquad
\theta _{{\bar {v}}{\bar {v}}}  =
- \frac {1}{12\pi } {\bar {C}}^{\frac {1}{2}} \left( {\bar {C}}^{-\frac {1}{2}} \right) _{,{\bar {v}}{\bar {v}}}
\qquad
\theta _{{\bar {u}}{\bar {v}}}  =  \theta _{{\bar {v}}{\bar {u}}} =0.
\end{equation}
The third term in (\ref{eq:DFU}) is the usual conformal trace anomaly.

\section{Collapse scenario}
\label{sec:collapse}

\begin{figure}\sidecaption
\includegraphics[width=6cm]{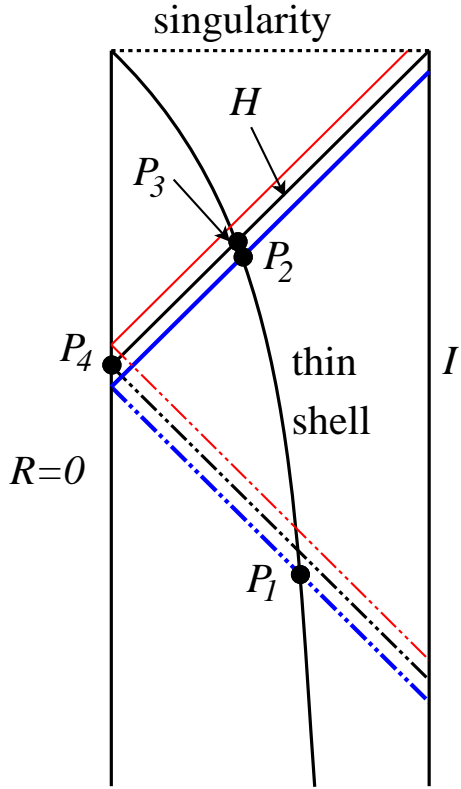}
\caption{Space-time diagram for a black hole formed by gravitational collapse of a thin shell in anti-de Sitter space-time.
The black curve is the collapsing shell; null infinity is denoted ${\mathcal {I}}$ and ${\mathcal {H}}$ is the event horizon. The horizontal dotted line at the top of the diagram is the space-time singularity.
The point $P_{3}$ is where the collapsing shell crosses the event horizon.
The black dot-dashed curve is an ingoing null ray which is reflected at the origin at the point $P_{4}$ and then forms the event horizon.
Shown in blue (thick lines) is a null ray which enters the shell just before the black ray, emerging just before the horizon forms. The blue ray enters the shell at the point $P_{1}$ and exits at the point $P_{2}$.  The red (thin) lines show a null ray which enters the shell just after the black ray, and hence remains inside the horizon.  Solid blue/red (thick/thin) lines denote outgoing null rays, dot-dashed lines ingoing null rays.}
\label{fig:one}
\end{figure}

We consider a thin shell of matter collapsing in two-dimensional anti-de Sitter (adS) space-time, shown in figure \ref{fig:one}.
The shell has radius $S(t)$ at time $t$.
In the far past, the shell is static and has radius $S_{0}$. It begins to collapse at some time $t=t_{0}$ and the collapse continues until a black hole of radius $r_{h}$ is formed.
We assume that $S_{0}$ is much larger than $r_{h}$.

Exterior to the shell, the space-time has the Schwarzschild-anti-de Sitter (SadS) metric
\begin{equation}
ds^{2}_{E} = -\left( 1 - \frac{2M}{r} + \frac {r^{2}}{\ell ^{2}} \right) dt^{2} + \left( 1 - \frac {2M}{r} + \frac {r^{2}}{\ell ^{2}} \right) ^{-1}
dr^{2},
\label{eq:metricSadS}
\end{equation}
in terms of static coordinates $(t,r)$, where $\ell $ is the adS radius.
Inside the shell, the space-time is pure adS with metric
\begin{equation}
ds^{2}_{I} = - \left( 1 + \frac {R^{2}}{\ell ^{2}} \right) dT^{2} + \left( 1 + \frac {R^{2}}{\ell ^{2}} \right) ^{-1} dR^{2} ,
\label{eq:metricadS}
\end{equation}
again in terms of static coordinates, in this case denoted $(T, R)$.
On the shell, at $r=S(t)=R$, the exterior (\ref{eq:metricSadS}) and interior (\ref{eq:metricadS}) metrics must match, which leads to the condition
\begin{eqnarray}
& &  -\left( 1 - \frac {2M}{S} + \frac {S^{2}}{\ell ^{2}} \right) \left( \frac {dt}{dT} \right) ^{2}
+ \left( 1 - \frac {2M}{S} + \frac {S^{2}}{\ell ^{2}} \right) ^{-1} \left( \frac {dS}{dT}  \right) ^{2}
\nonumber \\
& & \qquad \qquad \qquad =
-\left( 1 + \frac {S^{2}}{\ell ^{2}} \right) + \left( 1 + \frac {S^{2}}{\ell ^{2}} \right) ^{-1} \left( \frac {dS}{dT} \right) ^{2} .
\label{eq:shell}
\end{eqnarray}
Exterior to the shell, we define the usual ``tortoise'' coordinate $r_{*}$ by
\begin{equation}
\frac {dr_{*}}{dr} = \left( 1 - \frac {2M}{r} + \frac {r^{2}}{\ell ^{2}} \right) ^{-1}.
\label{eq:rstarext}
\end{equation}
By an appropriate choice of the constant of integration, we may take $r_{*}=0$ at ${\mathcal {I}}$, where $r\rightarrow \infty $.
Before the formation of the event horizon, $r_{*}$ is defined for all $r>S(t)$.
Once the event horizon has formed, $r_{*}\rightarrow -\infty $ as $r\rightarrow r_{h}$.
We then define null coordinates $\left( u, v \right)$ in the region exterior to the shell or event horizon by
\begin{equation}
u = t - r_{*}, \qquad v = t+ r_{*},
\label{eq:extnull}
\end{equation}
where $u$ is constant on outgoing rays and $v$ is constant on ingoing rays.
Interior to the shell, we also define a ``tortoise'' coordinate $R_{*}$ by
\begin{equation}
\frac {dR_{*}}{dR} = \left( 1 + \frac {R^{2}}{\ell ^{2}} \right) ^{-1},
\label{eq:rstarint}
\end{equation}
so that, with a suitable choice of integration constant,
\begin{equation}
R_{*} = \frac {1}{\ell } \tan ^{-1} \left( \frac {R}{\ell } \right) ,
\end{equation}
and $R_{*}=0$ at the origin $R=0$.
Null coordinates $(U,V)$ inside the shell are then defined by
\begin{equation}
U = T - R_{*}, \qquad V = T + R_{*},
\label{eq:intnull}
\end{equation}
where $U$ is constant on outgoing rays and $V$ is constant on ingoing rays.

Since adS is not a globally hyperbolic space-time, we need to impose boundary conditions on the scalar field $\phi $.
We use reflective boundary conditions \cite{Avis:1977yn}, setting the scalar field to vanish on ${\mathcal {I}}$.
Exterior to the shell, the scalar field modes therefore take the form
\begin{equation}
\phi _{\omega } = \frac {1}{{\sqrt {4\pi \left|\omega  \right| }}} \left( e^{-i\omega u} - e^{-i\omega v} \right) .
\label{eq:extmodes}
\end{equation}
Inside the shell, as can be seen in figure \ref{fig:one}, there is an additional time-like boundary at $R=0$. We also use Dirichlet boundary conditions here, setting $\phi = 0$ at $R=0$,
so that the field modes have the following form inside the shell:
\begin{equation}
\Phi _{\omega } = \frac {1}{\sqrt {4\pi \left| \omega \right| }} \left( e^{-i\omega U}- e^{-i\omega V } \right) .
\label{eq:intmodes}
\end{equation}

At very early times, before the shell has started to collapse, the space-time is static.
Exterior to the shell, the massless scalar field modes take the form (\ref{eq:extmodes}).
In this region, we may take the null coordinates $({\bar {u}},{\bar {v}})$ which define the ingoing and outgoing waves to be
\begin{equation}
{\bar {u}} =u , \qquad {\bar {v}} = v.
\label{eq:earlyuv}
\end{equation}
We define positive frequency modes to have $\omega >0$, and expand the scalar field in terms of the modes (\ref{eq:extmodes}):
\begin{equation}
\phi = \sum _{\omega } a_{\omega } \phi _{\omega } + a_{\omega }^{\dagger } \phi _{\omega }^{*},
\end{equation}
where a star denotes complex conjugate.
The field is quantized by promoting the expansion coefficients $(a_{\omega }, a_{\omega }^{\dagger })$ to operators and then the vacuum state of interest $| 0 \rangle $ is annihilated by the $a_{\omega }$ operators, namely $a_{\omega } | 0 \rangle =0$.

At late times, long after the black hole has formed, the scalar field modes exterior to the event horizon again have the form (\ref{eq:extmodes}).
However, the coordinates $({\bar {u}},{\bar {v}})$ will {\em {not}} be given by (\ref{eq:earlyuv}) in this region, because the space-time is dynamical.
Let ${\bar {u}}=f(u)$ be a function of the exterior outgoing null coordinate.
Without loss of generality we may assume that ${\bar {u}}={\bar {v}}$ at ${\mathcal {I}}$.
Therefore, at late times, the field modes have the following form in $({\bar {u}}, {\bar {v}})$ coordinates:
\begin{equation}
\phi \propto e^{-i\omega {\bar {u}}} - e^{-i\omega {\bar {v}}}
\label{eq:latephi}
\end{equation}
in order that the field vanishes on ${\mathcal {I}}$.
At late times, since ${\bar {u}}=f(u)$, the form (\ref{eq:latephi}) becomes
\begin{equation}
\phi \propto e^{-i\omega f(u)} - e^{-i\omega {\bar {v}}}.
\end{equation}
Since on ${\mathcal {I}}$ the exterior null coordinates are equal, $u=v$, it must be the case that
${\bar {v}} = f(v)$ at late times. In other words, the null coordinates $({\bar {u}}, {\bar {v}})$ have the same functional form at late times in terms of the exterior null coordinates $(u,v)$.

To find this functional form, consider modes seen by an observer far from the black hole at late times.
These modes must have exited the collapsing shell shortly before the shell crossed the event horizon.
Therefore consider the outgoing null ray shown in blue (thick line) in figure \ref{fig:one}.
This ray exits the shell at the point $P_{2}$ just before the horizon forms.
Let the finite time (as seen by observers inside the shell) at which the shell enters the event horizon (the point $P_{3}$ in figure \ref{fig:one}) be $T=T_{0}$.
In a neighbourhood of the point $P_{3}$,
we approximate the path of the shell as a function of the interior coordinate time $T$ as:
\begin{equation}
S(T) \approx r_{h} + {\mathcal {A}} \left( T_{0} - T \right) ,
\label{eq:p2approx}
\end{equation}
where ${\mathcal {A}}$ is a constant and we have ignored terms of higher order in $T_{0}-T$.
Substituting (\ref{eq:p2approx}) in (\ref{eq:shell}), we find, in a neighbourhood of $P_{3}$, that
\begin{equation}
\left( \frac {dt}{dT} \right) ^{2} \approx h(r_{h})^{-2} \left( T_{0} - T \right) ^{-2},
\end{equation}
where we have retained only the leading order term.
The function $h(r)$ is defined by
\begin{equation}
1 - \frac {2M}{r} + \frac {r^{2}}{\ell ^{2}} = \left( r - r_{h} \right) h(r),
\end{equation}
so that $h(r_{h})$ is finite.
Therefore, at $P_{2}$, the interior and exterior time coordinates are related by
\begin{equation}
t \approx - h(r_{h})^{-1} \ln \left( \frac {T_{0}-T}{{\mathcal {B}}} \right) ,
\end{equation}
for some constant ${\mathcal {B}}$, where we have ignored terms which are finite as $T\rightarrow T_{0}$.
Similarly, in a neighbourhood of $P_{3}$, the equation governing the exterior tortoise coordinate $r_{*}$ (\ref{eq:rstarext}) takes the form
\begin{equation}
\frac {dr_{*}}{dr} \approx \frac {1}{h(r_{h})(r-r_{h})} ,
\end{equation}
where we have ignored terms which are finite as $r\rightarrow r_{h}$. Integrating, we find
\begin{equation}
r_{*} \approx h(r_{h})^{-1} \ln \left( \frac {S-r_{h}}{{\mathcal {B}}'} \right)
\approx h(r_{h})^{-1} \ln \left( \frac {{\mathcal {A}} \left( T_{0} - T \right) }{{\mathcal {B}}'} \right) ,
\end{equation}
where we have retained only the leading order terms and ${\mathcal {B}}'$ is another constant.
Therefore, at $P_{2}$, we have
\begin{equation}
u \approx -\frac {2}{h(r_{h})} \ln \left( \frac {T_{0}-T}{\mathcal {B}} \right) ,
\label{eq:u1}
\end{equation}
plus terms which are finite as $T\rightarrow T_{0}$.
Considering now the internal null coordinates, in our small neighbourhood of $P_{3}$, we have
\begin{eqnarray}
U = T- R_{*} & \approx & T - R_{*}(r_{h}) - \left. \frac {dR_{*}}{dr}\right|_{r=r_{h}} \left( S-r_{h} \right)
\nonumber \\
& \approx & T - \frac {1}{\ell } \tan ^{-1} \left( \frac {r_{h}}{\ell } \right) - {\mathcal {A}} \left( 1+ \frac {r_{h}^{2}}{\ell ^{2}}\right) ^{-1} \left( T_{0}-T \right) ,
\end{eqnarray}
ignoring higher-order terms.
Therefore, at $P_{2}$, the interior and exterior outgoing null coordinates are related by
\begin{equation}
u \approx - \frac {2}{h(r_{h})} \ln \left( \frac {U_{0}-U}{{\mathcal {C}}} \right) ,
\label{eq:U1}
\end{equation}
where $U_{0}$ is the value of the interior null coordinate $U$ at $P_{3}$ (which is finite), ${\mathcal {C}}$ is a constant and we have ignored
terms which are finite as $U\rightarrow U_{0}$.
Note that $U<U_{0}$ at $P_{2}$.
To match the modes (\ref{eq:extmodes}, \ref{eq:intmodes}) at $P_{2}$, such that the boundary conditions at both ${\mathcal {I}}$ and $R=0$ are satisfied,
it must be the case that the functional relationship between $v$ and $V$ is the same as that between $u$ and $U$ at $P_{2}$, namely:
\begin{equation}
v \approx -\frac {2}{h(r_{h})} \ln \left( \frac {U_{0}-V}{{\mathcal {C}}} \right) .
\label{eq:V1}
\end{equation}

We now need to trace the ray back to early times.  Following \cite{Unruh:1976db,Davies:1976ei}, we assume that the shell collapses sufficiently quickly that when the blue (thick line) ray in figure \ref{fig:one} enters the shell (at the point $P_{1}$), the collapse process has not started and the shell radius is much larger than the event horizon radius $r_{h}$.
In a neighbourhood of $P_{1}$, it is the case that $S-r_{h}$ is finite, $dS/dT$ can be taken to be approximately zero, and therefore, using (\ref{eq:shell}), $dt/dT$ can also be taken to be approximately constant.
Hence the exterior time coordinate $t$ is approximately a linear function of $T$, and similarly the radial coordinate
$r$ is approximately a linear function of $R$.
Therefore we can relate the interior and exterior null coordinates in a neighbourhood of $P_{1}$ by
\begin{equation}
U \approx a u + b, \qquad V \approx a v  + b,
\label{eq:p1rels}
\end{equation}
for some constants $a$ and $b$, whose values are not important.
At times earlier than the point $P_{1}$, the shell will again have a radius much larger than $r_{h}$, and so the interior and exterior null coordinates are approximately related by linear functions at each point on the shell at a time prior to $P_{1}$.

Combining the relations (\ref{eq:p1rels}) between the interior and exterior coordinates at early times and (\ref{eq:U1}, \ref{eq:V1}) at late times, we can now define the null coordinates $({\bar {u}}, {\bar {v}})$ at late times for the modes (\ref{eq:DFUmodes}), implicitly via the following relations:
\begin{equation}
u = -\frac {2}{h(r_{h})} \ln \left( \frac {{\bar {u}}_{0}-{\bar {u}}}{\mathcal {D}} \right) + {\mathcal {E}}({\bar {u}}),
\qquad
v = -\frac {2}{h(r_{h})} \ln \left( \frac {{\bar {u}}_{0}-{\bar {v}}}{\mathcal {D}} \right) + {\mathcal {E}}({\bar {v}}),
\label{eq:lateuv}
\end{equation}
where ${\mathcal {D}}$ is a constant and ${\mathcal {E}}({\bar {u}})$ is a function which is finite as ${\bar {u}}\rightarrow {\bar {u}}_{0}$, for a constant ${\bar {u}}_{0}$ which is the value of ${\bar {u}}$ on the future event horizon ${\mathcal {H}}$.

Modes on the space-time exterior to the horizon at late times must have ${\bar {u}}<{\bar {u}}_{0}$; those with ${\bar {u}}>{\bar {u}}_{0}$ lie behind the event horizon.
Similarly, there is a maximum value of ${\bar {v}}$, say ${\bar {v}}_{0}$, for modes exterior to the event horizon at late times.
This corresponds to the value of ${\bar {v}}$ for the ingoing ray which, after reflection at the origin at the point $P_{4}$ in figure \ref{fig:one}, forms the event horizon.
The field modes (\ref{eq:latephi}) must vanish at the origin, in particular at the point $P_{4}$ where ${\bar {u}}={\bar {u}}_{0}$ and ${\bar {v}}={\bar {v}}_{0}$.
Therefore we have ${\bar {u}}_{0} = {\bar {v}}_{0}$.

\section{$\langle T_{\mu \nu }\rangle $ for the collapse scenario}
\label{sec:Tmunu}

We now use the definitions (\ref{eq:earlyuv}, \ref{eq:lateuv}) to compute the stress-energy tensor in the vacuum state $| 0 \rangle $ using the method outlined in section \ref{sec:DFU}.
We consider only the space-time exterior to either the collapsing shell or the event horizon.
At early times, using (\ref{eq:metricSadS}, \ref{eq:earlyuv}), the metric (\ref{eq:metric}) is simply
\begin{equation}
ds^{2} = - \left( 1 - \frac {2M}{r} + \frac {r^{2}}{\ell ^{2}} \right) \, d{\bar {u}} \, d{\bar {v}},
\end{equation}
so that the conformal factor is
\begin{equation}
{\bar {C}}_{\rm {early}} ({\bar {u}}, {\bar {v}}) = C(r) = 1 - \frac {2M}{r} + \frac {r^{2}}{\ell ^{2}} ,
\label{eq:Cearly}
\end{equation}
where $C(r)$ is an implicit function of $(u, v)$.
From (\ref{eq:lateuv}), at late times we have
\begin{equation}
{\mathcal {F}}({\bar {u}}) = \frac {du}{d{\bar {u}}} = \frac {2}{h(r_{h}) \left( {\bar {u}}_{0} - {\bar {u}} \right) } + {\mathcal {E}}'({\bar {u}}),
\qquad
{\mathcal {F}}({\bar {v}})=\frac {dv}{d{\bar {v}}} = \frac {2}{h(r_{h}) \left( {\bar {u}}_{0} - {\bar {v}} \right) } + {\mathcal {E}}'({\bar {v}}).
\end{equation}
Therefore the metric (\ref{eq:metric}) takes the form
\begin{equation}
ds^{2} = - C(r) \frac {du}{d{\bar {u}}} \frac {dv}{d{\bar {v}}} \, d{\bar {u}} \, d{\bar {v}},
\end{equation}
and the conformal factor is
\begin{equation}
{\bar {C}}_{\rm {late}} ({\bar {u}}, {\bar {v}} ) = C(r) {\mathcal {F}}({\bar {u}}) {\mathcal {F}}({\bar {v}}) .
\label{eq:Clate}
\end{equation}

We are interested in the difference in the expectation value of the stress-energy tensor $\langle T_{\mu \nu }\rangle $ (\ref{eq:DFU}) at early and late times.
As explained in section \ref{sec:DFU}, the first term in (\ref{eq:DFU}) depends on the ranges of the coordinates $({\bar {u}}, {\bar {v}})$ and hence is the same at early and late times (since we have the same set of field modes at both early and late times).
The third term in (\ref{eq:DFU}) is also the same at both early and late times, so, following \cite{Davies:1976ei,Tadaki:1990cg} we focus on the second term, ${}^{2}T_{\mu \nu } =\theta _{\mu \nu }$.
Consider $\theta _{{\bar {u}}{\bar {u}}}$; the calculation of $\theta _{{\bar {v}}{\bar {v}}}$ will proceed similarly.

At early times, using (\ref{eq:earlyuv}, \ref{eq:Cearly}), we have simply
\begin{equation}
\theta _{{\bar {u}}{\bar {u}}}^{\rm {early}} =
-\frac {1}{12\pi } {\sqrt {{\bar {C}}_{\rm {early}}}} \left( \frac {1}{{\sqrt {{\bar {C}}_{\rm {early}}}}} \right) _{,{\bar {u}}{\bar {u}}}
=-\frac {1}{12\pi } C^{\frac {1}{2}} \left( C^{-\frac {1}{2}} \right) _{,uu}
\end{equation}
and hence
\begin{equation}
{}^{2}T^{{\rm {early}}}_{uu} = {}^{2}T^{{\rm {early}}}_{{\bar {u}}{\bar {u}}} = -\frac {1}{12\pi } C^{\frac {1}{2}} \left( C^{-\frac {1}{2}} \right) _{,uu}.
\end{equation}

The calculation for late times is more complicated because of the relationship (\ref{eq:lateuv}) between the null coordinates $(u,v)$ and $({\bar {u}}, {\bar {v}})$.
Using (\ref{eq:Clate}) we have
\begin{eqnarray}
\theta _{{\bar {u}}{\bar {u}}}^{\rm {late}} & = &
-\frac {1}{12\pi } {\sqrt {{\bar {C}}_{\rm {late}}}} \left( \frac {1}{{\sqrt {{\bar {C}}_{\rm {late}}}}} \right) _{,{\bar {u}}{\bar {u}}}
\nonumber \\ & = &
-\frac {1}{12\pi } \left\{
{\mathcal {F}}({\bar {u}})^{2} C^{\frac {1}{2}} \left( C^{-\frac {1}{2}} \right) _{,uu} + \frac {3}{4{\mathcal {F}}({\bar {u}})^{2}} \left[
{\mathcal {F}}'({\bar {u}}) \right] ^{2} - \frac {1}{2{\mathcal {F}}({\bar {u}})} {\mathcal {F}}''({\bar {u}})
\right\} .
\nonumber \\ & &
\end{eqnarray}
To compare the stress-energy tensors at early and late times as seen by a static observer far from the black hole, we require the component ${}^{2}T_{uu}^{{\rm {late}}}$ in the original exterior null coordinates $(u,v)$. This is given by
\begin{eqnarray}
{}^{2}T_{uu}^{{\rm {late}}} & = &
\left[ {\mathcal {F}}({\bar {u}}) \right] ^{-2} \theta _{{\bar {u}}{\bar {u}}}^{\rm {late}}
\nonumber \\
 & = & -\frac {1}{12\pi } \left\{
C^{\frac {1}{2}} \left( C^{-\frac {1}{2}} \right) _{,uu} + \frac {3}{4{\mathcal {F}}({\bar {u}})^{4}} \left[
{\mathcal {F}}'({\bar {u}}) \right] ^{2} - \frac {1}{2{\mathcal {F}}({\bar {u}})^{3}} {\mathcal {F}}''({\bar {u}})
\right\} .
\end{eqnarray}
Therefore the difference in the component $\langle T_{uu} \rangle $ of the stress-energy tensor between early and late times is
\begin{eqnarray}
{}^{2}T_{uu}^{{\rm {late}}} - {}^{2} T_{uu}^{{\rm {early}}}
 & = &
-\frac {1}{12\pi }\left\{ \frac {3}{4{\mathcal {F}}({\bar {u}})^{4}} \left[
{\mathcal {F}}'({\bar {u}}) \right] ^{2} - \frac {1}{2{\mathcal {F}}({\bar {u}})^{3}} {\mathcal {F}}''({\bar {u}})
\right\}
\nonumber \\
 & = & \frac {h(r_{h})^{2}}{192\pi } =\frac {\pi T_{H}^{2}}{12},
 \label{eq:Tuu}
\end{eqnarray}
where we have ignored terms which vanish as ${\bar {u}} \rightarrow {\bar {u}}_{0}$ for late times, and
\begin{equation}
T_{H} = \frac {h(r_{h})}{4\pi }
\label{eq:Th}
\end{equation}
is the Hawking temperature of the SadS black hole.
The calculation of the difference in the component $\langle T_{vv} \rangle $ between early and late times proceeds in a similar way and gives the same answer:
\begin{equation}
{}^{2}T_{vv}^{{\rm {late}}} - {}^{2} T_{vv}^{{\rm {early}}}
= \frac {\pi T_{H}^{2}}{12},
\label{eq:Tvv}
\end{equation}
again ignoring terms which vanish as ${\bar {v}} \rightarrow {\bar {v}}_{0}={\bar {u}}_{0}$ for late times.

The final expression (\ref{eq:Tuu}) for the difference in the component $T_{uu}$ between late and early times is the same as that found in \cite{Davies:1976ei} for an asymptotically flat Schwarzschild black hole formed by gravitational collapse.  However, the difference in the other component (\ref{eq:Tvv}) vanishes in the asymptotically flat case \cite{Davies:1976ei}.  For an asymptotically flat black hole formed by gravitational collapse, at late times there is an outgoing flux of thermal particles \cite{Davies:1976ei}. In the present case, the outgoing flux in the $T_{uu}$ component is matched by an incoming flux in the $T_{vv}$ component.
We conclude that at late times the black hole is in thermal equilibrium with a heat bath at the Hawking temperature.

\section{Conclusions}
\label{sec:conc}

In this paper we have considered a massless quantum scalar field on a two-dimensional space-time describing a shell of matter collapsing to form a black hole in anti-de Sitter (adS) space-time.  Applying the techniques of quantum field theory in curved space, the background geometry is purely classical and we ignore the backreaction of the quantum field on the space-time geometry.
At early times, long before the shell starts to collapse, the scalar field exterior to the shell is in the vacuum state, which is the state analogous to the Boulware \cite{Boulware:1974dm} vacuum on an eternal black hole space-time.
We apply the method of Davies-Fulling-Unruh (DFU) \cite{Davies:1976ei} to compute the difference in the renormalized expectation value of the stress-energy tensor, $\langle T_{\mu \nu } \rangle $, between early and late times.
At late times, we find that the Schwarzschild-adS (SadS) black hole, formed by the gravitational collapse of the shell, is in thermal equilibrium with a heat bath at the Hawking temperature.
Therefore the quantum state at late times is analogous to the Hartle-Hawking state \cite{Hartle:1976tp} on an eternal black hole geometry.
We have studied a very simple two-dimensional toy model, ignoring the backreaction of the quantum field on the space-time geometry, and the scattering effects which are important for black holes in four or more space-time dimensions.

In the introduction, we asked whether there is a state for SadS black hole space-times which is analogous to the Unruh vacuum \cite{Unruh:1976db} for asymptotically flat black holes.  Our analysis has indicated that the state at late times for a quantum field on a black hole in adS formed by gravitational collapse is in fact the Hartle-Hawking vacuum.
It should be emphasised however, that our choice of reflective boundary conditions on the adS boundary is crucial for obtaining this result.
For example, a toy model very similar to ours is studied in \cite{Hemming:2000as}, where it is concluded that the state soon after the black hole forms will be analogous to the asymptotically flat space-time Unruh state (that is, an outgoing flux of thermal radiation but no incoming flux).  However,
in \cite{Hemming:2000as} reflecting boundary conditions are not imposed at ${\mathcal {I}}$.
Another key part of our analysis is that we have assumed that the shell collapses rapidly to form the black hole. In \cite{Giddings:2001ii}, it was shown that the quantum field remains in the Boulware-like state if the shell is lowered quasi-statically towards the event horizon.

Our analysis makes no distinction between ``small'' and ``large'' (relative to the adS length scale $\ell $) black holes, the latter of which are thermodynamically stable \cite{Hawking:1982dh} while the former are thermodynamically unstable and are therefore expected to evaporate.
This is because we have considered a fixed (but dynamical) space-time geometry and ignored backreaction effects, which may be very significant for a collapsing shell of quantum matter in adS (see, for example, \cite{Alberghi:2003pr}).
It is known that even ``large'' asymptotically adS black holes may evaporate if the boundary conditions on ${\mathcal {I}}$ are changed, for example, by coupling to an external field so that the adS boundary is at least partially absorbing (see, for example, \cite{Rocha:2008fe,Almheiri:2013hfa,VanRaamsdonk:2013sza}).
It would be interesting to extend our work here to different boundary conditions on ${\mathcal {I}}$, but
we leave this question for a future investigation.

\begin{acknowledgements}
P.G.A.~thanks Jorma Louko for helpful discussions. E.W.~thanks Gavin Duffy for helpful discussions.
The work of E.W.~is supported by the Lancaster-Manchester-Sheffield Consortium for Fundamental Physics under STFC grant ST/L000520/1.
\end{acknowledgements}

\end{document}